\documentclass[journal,comsoc]{IEEEtran}

\usepackage{xcolor}
\definecolor{mydarkgreen}{RGB}{0, 192, 0} 
\usepackage{hyperref}
\hypersetup{colorlinks,linkcolor={blue},citecolor={blue},urlcolor={mydarkgreen}}
\usepackage{cite}
\usepackage{amsmath,amssymb,amsfonts}
\usepackage{mathtools, cuted}   
\usepackage{pifont}

\usepackage{algorithm}
\usepackage{algorithmic}
\usepackage{graphicx}
\usepackage{textcomp}
\usepackage{orcidlink}
\usepackage{bm} 
\def\BibTeX{{\rm B\kern-.05em{\sc i\kern-.025em b}\kern-.08em
    T\kern-.1667em\lower.7ex\hbox{E}\kern-.125emX}}

\usepackage{array}
\usepackage[caption=false,font=normalsize,labelfont=sf,textfont=sf]{subfig}
\usepackage{url}
\usepackage{verbatim}
\usepackage{booktabs}

\begin{document}

\title{Leveraging Persistent Field-of-View and Multipath Segment Routing for Traffic Engineering in LEO Satellite Constellations}

\author{Kai~Wang \orcidlink{0000-0002-6170-4744} and Chee~Wei~Tan \orcidlink{0000-0002-6624-9752}
\thanks{This research was supported in part by the Singapore Ministry of Education Academic Research Fund Tier 2 MOE-T2EP20224-0009 and NTU startup \emph{(Corresponding author: Chee Wei Tan)}.}
\thanks{The authors Kai Wang and Chee Wei Tan are with the College of Computing and Data Science (CCDS), Nanyang Technological University, Singapore 639798 (e-mail: wang1899@e.ntu.edu.sg; cheewei.tan@ntu.edu.sg).}
}

\markboth{IEEE Transactions on Network Science and Engineering,~Vol.~14, No.~8, August~2026}%
{Kai Wang \MakeLowercase{\textit{et al.}}: Leveraging Persistent Field-of-View and Multipath Segment Routing for Traffic Engineering in LEO Satellite Constellations}


\maketitle

\begin{abstract}
The emergence of low earth orbit (LEO) satellite constellations poses fundamental challenges for traditional routing paradigms due to high node mobility, frequent topology changes, and constrained on-board resources. This paper proposes a segment routing (SR) framework that offloads route computation and control signaling from resource-constrained LEO switches to more powerful medium earth orbit (MEO) controllers. We first establish sufficient conditions for the persistence of user-to-satellite and LEO-to-MEO field-of-view (FoV) connectivity throughout a virtual topology (VT) snapshot interval, enabling handover-free operation. Leveraging this persistence, we formulate the MEO control domain partitioning problem as a mixed-integer linear program (MILP) to minimize control overhead while assigning disjoint segment routing domains (SRDs) to MEO controllers. Within each SRD, we develop a multipath SR-based traffic engineering (SR-TE) formulation that splits traffic among four minimum-hop-count path types to minimize maximum link utilization. We solve this linear program (LP) using a distributed primal-dual hybrid gradient (PDHG) approach that preserves SRD-level privacy and requires only AllReduce communication. Simulations on the Starlink/GPS combination demonstrate our approach reduces maximum link utilization by more than 13.09\%, 11.84\%, and 12.50\%, respectively, compared to traditional minimum hop count routing (MHCR) and traffic-oblivious load-balancing methods for Starlink, Kuiper, and OneWeb.
\end{abstract}


\section{Introduction}\label{sec:introduction}

\IEEEPARstart{T}{he} rapid proliferation of low earth orbit (LEO) constellations, such as Starlink \cite{starlink}, Kuiper \cite{kuiper}, Telesat \cite{telesat}, and OneWeb \cite{oneweb}, has ushered in a new era of global low-latency broadband connectivity. However, the inherent characteristics of these networks with high node mobility, frequent topology changes, and constrained on-board resources pose fundamental challenges for traditional routing paradigms. To mitigate route flapping and convergence delays that result from the dynamic nature, the maintenance of LEO constellations involves intense control message exchange. However, inter-satellite link (ISL) instability forces more frequent control message retransmissions \cite{isl1,isl2}, making larger domains inefficient due to the number of unstable links and cascading failures, so smaller, distributed domains help localize retransmissions and reduce overall control overhead. In addition to more control overhead and less responsiveness, the centralized authority makes the network vulnerable to central software failures, as demonstrated by the notorious 2025 outage \cite{outage}.

Fortunately, medium earth orbit (MEO) satellites, operating at altitudes between 8,000 and 20,000 km, offer a wide-area and stable recipe for managing LEO satellite networks with longer orbital periods and wider geographical coverage. Typically, the MEO constellation (e.g., 6-12 satellites at the minimum) provides global coverage while keeping the number of satellites low. This unique vantage point suggests a natural architectural hierarchy subject to the physical line-of-sight (LoS) and field-of-view (FoV) \cite{fov1} constraints. At any time, in addition to LoS visibility, the FOV constraint (e.g., 25-40\textdegree\ minimum elevation angle \cite{fov2}) must be respected. Using the virtual topology (VT) strategy, we partition the dynamic constellation into a sequence of static topology snapshots. For each snapshot, a mixed-integer linear program (MILP) partitions the LEO constellation into disjoint control regions, where the optimal solution for a given snapshot may depend on that of the previous snapshot. The MEO controller is responsible for updating the states of the LEO constellation, running distributed path computation algorithms, and transmitting forwarding instructions downward to LEO satellites. Critically, multiple MEO satellites may have overlapping FoVs, providing redundancy and enabling seamless handoff as LEO satellites migrate from one MEO's control region to another.

Both the MEO and LEO layers evolve deterministically over time, as do the FoVs between them. Since MEO nodes orbit at a significantly higher altitude, their FoVs with LEO nodes maintain sufficient width and stability over extended periods, making MEO satellites suited to serve as regional controllers during each VT snapshot. Similarly, with careful assignment of terrestrial users to LEO satellites within LoS and FoV constraints, stable ground-to-satellite link (GSL) availability can be achieved, enabling fixed user-LEO-MEO associations. Based on these associations, we formulate an LP to optimize TE in the LEO layer. While incumbent LP solvers leveraging second-order statistics offer fast convergence, they incur substantial communication overhead that renders them impractical for distributed solvers. At the start of each VT snapshot, we can obtain a solution using centralized LP solvers; however, real-time traffic fluctuations require LEO satellites to measure or forecast \cite{forecast} demand and fine-tune the solution accordingly, which can be initialized from the current solution to avoid a cold start. This motivates the use of distributed optimization methods, typically exchanging variables among nodes at significant communication cost. To address this challenge, we leverage D-PDLP \cite{d-pdlp}, a communication-efficient distributed implementation of the primal-dual hybrid gradient (PDHG) algorithm \cite{pdhg} for massive-scale LP problems, which has recently been applied to satellite communications \cite{pdhg2,pdhg3}. Specifically, we leverage D-PDLP to distribute routing computations across MEO satellites while minimizing exchanged variable messages, solving the routing LP with a distributed first-order algorithm that exploits multi-processor parallelism and requires less frequent communication than traditional methods such as ADMM \cite{admm}.

SRv6 (segment routing over IPv6) \cite{sr} is widely regarded as a promising solution for LEO constellation networks. Within a given MEO controller's domain, the LEO satellites naturally form a self-contained segment routing domain (SRD) \cite{ToN17} that operates entirely in the data plane. For load balancing, we solve a segment routing-based traffic engineering (SR-TE) problem. Then, we translate the resulting routing scheme into feasible segment lists (SLs) that respect the constraints of ternary content addressable memory (TCAM) hardware.

In summary, this paper leverages a hierarchical design philosophy and proposes a novel communication-efficient load-balancing framework, which exploits the MEO-LEO hierarchy through a clean separation of control and data plane functions. Specifically, resource-constrained LEO satellites do not perform any route computation, topology discovery, or control signaling; these tasks are offloaded to the more powerful MEO layer. Our main contributions are as follows.

\begin{enumerate}
\item We present a practical framework of SR-based TE over a hierarchical MEO-LEO satellite network with explicit FoV connectivity constraints.
\item We propose a method to assign LEO satellite switches to MEO satellite controllers without handovers in the VT snapshot, pose the control domain partitioning problem, and formulate the problem as an MILP to optimize control overhead.
\item We propose a method to associate users to LEO satellites without handovers in the VT snapshot, formulate the SR-based TE as an LP to obtain an initial solution, and leverage D-PDLP, a communication-efficient distributed LP solver, for subsequent distributed adjustment of the solution in response to traffic demand fluctuations.
\item We evaluate HR-SR through simulation. Using realistic Walker-Delta LEO constellations (1,584 satellites at 550 km altitude) and a 24-satellite MEO constellation (20,200 km altitude), we demonstrate that our approach reduces maximum link utilization by more than 13.09\%, 11.84\%, and 12.50\% compared to traditional minimum hop count routing (MHCR) and traffic-oblivious load-balancing methods.
\end{enumerate}

The remainder of the paper is organized as follows. Section \ref{sec:satellitedynamics} presents fundamentals of satellite dynamics and constellations. Section \ref{sec:sr-background} provides preliminaries on the SR architecture and the SRD. Section \ref{sec:systemoverview} describes modular components of the SR framework over LEO constellations in a top-down approach. Section \ref{sec:lpformulation} formulates the MILP-based SRD partitioning problem. Section \ref{sec:sr-te} formulates the load balancing TE problem as an LP and proposes to solve with D-PDLP. Section \ref{sec:evaluation} introduces state-of-the-art benchmarks and compares their performance with the proposed method. Section \ref{sec:relatedworks} briefly surveys works related to SR optimization in LEO satellite networks, etc. To conclude, Section \ref{sec:conclusion} draws conclusions from both theory and experiments and outlooks future research possibilities.

\section{Satellite Dynamics} \label{sec:satellitedynamics}

To facilitate the ensuing discussion, this section presents the fundamentals of satellite dynamics, with most of the material adapted from literature \cite{dynamics1,dynamics2}. We list key notations used in this section in Table \ref{tab:satellitedynamics}.

\begin{table}[ht]
    \centering
    \caption{Notations Used in Satellite Dynamics}
    \label{tab:satellitedynamics}
    \begin{tabular}{m{2.0cm} m{6.0cm}}
        \hline
        \textbf{symbol} & \textbf{description} \\
        \vspace{1pt}
        $a>0$ & semi-major axis of orbit \\
        \vspace{1pt}
        $e>0$ & eccentricity of orbit \\
        \vspace{1pt}
        $\alpha\!\in\![0,\pi]$ & inclination of orbit \\
        \vspace{1pt}
        $\Omega\!\in\![-\pi,\pi)$ & right ascension of the ascending node \\
        \vspace{1pt}
        $\omega\!\in\![-\pi,\pi)$ & pariapsis of orbit \\
        \vspace{1pt}
        $\nu\!\in\![-\pi,\pi)$ & true anomaly of satellite \\
        \vspace{1pt}
        $r_e$ & radius of the Earth \\
        \vspace{1pt}
        $\phi$ & latitude on the sphere \\
        \vspace{1pt}
        $\lambda$ & longitude on the sphere \\
        \vspace{1pt}
        $h$ & altitude above the sphere \\
        \vspace{1pt}
        $\zeta(\nu)$ & longitude difference of satellite at true anomaly $\nu$ to its ascending node \\
        \vspace{1pt}
        $\psi\!\in\![0,\pi]$ & central angle between two points on the sphere \\
        \vspace{1pt}
        $\theta\!\in\![-\pi/2,\pi/2]$ & elevation angle from terrestrial user to satellite \\
        \vspace{1pt}
        $s$ & slant range between terrestrial user and satellite \\
        \vspace{1pt}
        $p, P$ & index of orbit, number of orbits \\
        \vspace{1pt}
        $q, Q$ & index of satellite, number of satellites per orbit \\
        \vspace{1pt}
        $F$ & phasing factor \\
        \vspace{1pt}
        $\Delta\Omega$ & RAAN difference between adjacent orbits \\
        \vspace{1pt}
        $\Delta\nu$ & phase offset between intra-orbit satellites \\
        \vspace{1pt}
        $\Delta f$ & difference in true anomaly between inter-orbit satellites \\
        \vspace{1pt}
        $\omega_e,\omega_s$ & angular speed of the Earth, satellite \\
        \vspace{1pt}
        $H_p,H_q,H$ & inter-plane, intra-plane, and total hop count \\
        \hline
    \end{tabular}
\end{table}

\subsection{Satellite Keplerian Elements}

An ascending node refers to the crossover of the northward satellite orbit and the Equatorial plane. With the ascending node, it is convenient to describe a satellite's current position with the six Keplerian elements. The semi-major axis $a$ and the eccentricity $e$ describe the size and shape of the elliptical orbit. The inclination $\alpha\!\in\![0,\pi]$, defined as the angle between the orbital plane and the equatorial plane at the ascending node, provides a quantitative measure of the orbit's tilt relative to the Equator. The right ascension of the ascending node (RAAN) $\Omega\!\in\![-\pi,\pi)$ is the angle between the ascending node and the Vernal Equinox, an Earth-centered inertial (ECI) reference frame centered on the Sun. The periapsis $\omega\!\in\![-\pi,\pi)$ is the angle within the orbital plane between the ascending node and the closest point of the Earth. The true anomaly $\nu\!\in\![-\pi,\pi)$ denotes the angle within the orbital plane between the satellite and the periapsis.

For ideal circular orbits where $e=0$, the semi-major axis equals the Earth's radius $r_e=6378.137$ km plus the satellite's altitude $h$ above the Earth's surface, i.e., $a=r_e+h$. However, the periapsis $\omega$ is not uniquely defined and we assume it to coincide with the ascending node, i.e., $\omega=0$. Consequently, the true anomaly $\nu$ is the angle between the satellite and the ascending node, which describes the angular position within its orbit. A satellite is either ascending or descending, according to its current true anomaly $\nu$. An ascending satellite rotates in the north-east direction and has $\nu\in[-\pi/2,\pi/2)$, while a descending satellite rotates in the south-west direction and has $\nu\in[-\pi,-\pi/2)\cup[\pi/2,\pi)$. In short, an circular and geocentric orbit is fully characterized by the radius $r_e$, inclination $\alpha$, and RAAN $\Omega$, and a satellite is specified by the true anomaly $\nu$.

\subsection{Satellite Geodetic Coordinates}

\begin{figure}[t]
\centering
\includegraphics[width=0.48\textwidth]{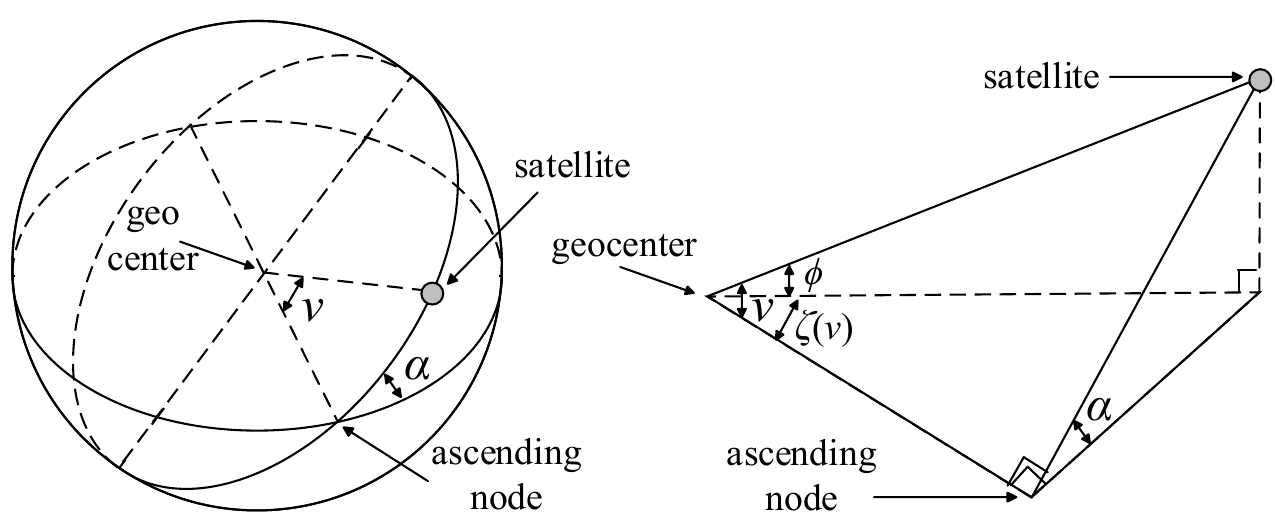}
\vspace{-5pt}
\caption{The Keplerian-to-geodetic conversion. In the right subplot, the triangle containing the inclination $\alpha$ lies in the tangential plane at the ascending node.}
\vspace{-10pt}
\label{fig:geodetic}
\end{figure}

Although the Keplerian elements fully specify an orbit and the position of an individual satellite, it is more beneficial to convert to geodetic coordinates for computational purposes. The geodetic coordinates, which are inherently Earth-fixed and rotate with the Earth, consist of the latitude $\phi$, the longitude $\lambda$, and the altitude $h$. At time $t$, the conversion formulas are
\begin{align}
    \phi_t=&\ \arcsin(\sin\alpha\cdot\sin\nu_t),\ \phi\in[-\alpha,\alpha],\label{eq:phi}\\
    \lambda_t=&\ \mathcal{N}(\Omega_t+\zeta(\nu_t)),\ \lambda\in[-\pi,\pi),\label{eq:lambda}\\
    h_t=&\ h,\label{eq:h}
\end{align}
where $\Omega_t=\Omega_0-\omega_e\cdot t$ is the longitude of the ascending node at time $t$, $\omega_e$ is the Earth's angular speed, $\nu_t=\nu_0+\omega_s\cdot t$ is the satellite's true anomaly at time $t$, $\omega_s$ is the satellite's angular speed at altitude $h$, and $\zeta(\nu_t)$ is the longitudinal difference of a satellite to its ascending node:
\begin{align}
    \zeta(\nu_t)=\arctan(\cos\alpha\cdot\tan\nu_t)+\begin{cases}
    0 & \text{asc. segment} \\
    \pi & \text{desc. segment}
\end{cases}.\label{eq:zeta}
\end{align}
We use the modulo function $\mathcal{N}(x)=((x+\pi)\text{ mod }2\pi)-\pi$ to constrain the angle within the domain of longitude. The relationships in Eqs. \eqref{eq:phi}-\eqref{eq:zeta} are visualized in Fig. \ref{fig:geodetic}.


\subsection{FoV Conditions for Communication} \label{subsec:condition}

\begin{figure}[t]
\centering
\includegraphics[width=0.48\textwidth]{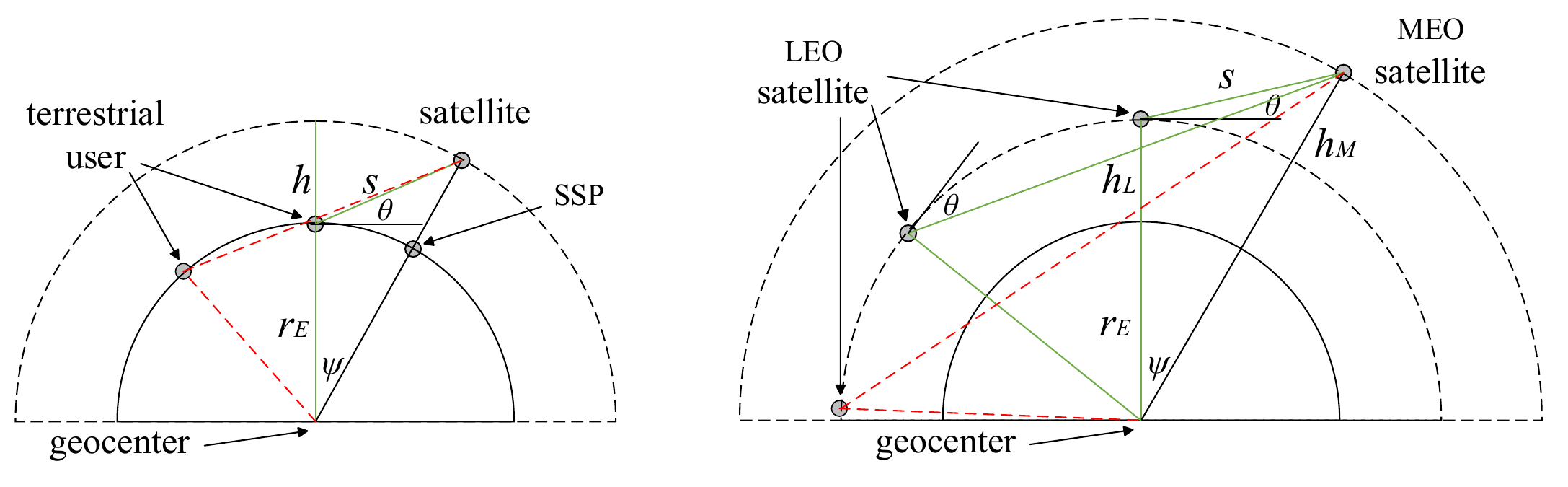}
\vspace{-5pt}
\caption{Illustration of LoS and FoV conditions with elevation angle $\theta$. Solid green lines satisfy LoS constraints (and also FoV constraints when $\theta\geq25^\circ$); dotted red lines violate LoS constraints.}
\vspace{-10pt}
\label{fig:elevation}
\end{figure}

At any time $t$, suppose that the user's geodetic coordinates are $(\phi_u,\lambda_u,0)$ and the satellite's geodetic coordinates are $(\phi_t, \lambda_t, h)$, given by Eqs. \eqref{eq:phi}-\eqref{eq:zeta}. The sub-satellite point (SSP), a.k.a. nadir or subastral point, is the perpendicular projection of the satellite onto the Earth's surface, whose latitude and longitude agree with the satellite's, and whose altitude equals 0 (i.e., on the Earth's surface). Given the longitudes and latitudes of the user and the SSP, we use the Haversine formula to calculate the central angle $\psi\in[0,\pi]$ between them:
\begin{align}
    \psi=&\ 2\arcsin\sqrt{\sin^2\big(\frac{\Delta\phi}{2}\big)+\cos{\phi_u}\cos{\phi_t}\sin^2\big(\frac{\Delta\lambda}{2}\big)},\label{eq:psi}
\end{align}
where $\Delta\phi=\phi_u-\phi_t$, $\Delta\lambda=\lambda_u-\lambda_t$. The LoS condition, i.e., the line joining the user and the satellite does not intersect the Earth's limb, is given by the Law of Sines:
\begin{align}
    \psi\leq\arccos{\frac{r_e}{r_e+h}}.\label{eq:los}
\end{align}
The user-to-satellite elevation angle $\theta\in[-\pi/2,\pi/2]$ is
\begin{align}
    \theta=\arctan\Big({\frac{\cos\psi-r_e/(r_e+h)}{\sin\psi}}\Big),\label{eq:theta}
\end{align}
and the constraint on the elevation angle requires \cite{fov1,fov2}
\begin{align}
    \theta\geq\theta_{us}=25\text{\textdegree}.\label{eq:fov}
\end{align}
For user-to-satellite communication via GSLs, we only require that Eq. \eqref{eq:fov} hold (that implies Eq. \eqref{eq:los}), which is equivalent to
\begin{align}\label{eq:fov1}
    {\frac{\cos\psi-r_e/(r_e+h)}{\sin\psi}}\geq\tan{\theta_{us}},\text{ i.e., }\psi\leq\psi_{us}.
\end{align}
For LEO-to-MEO communication via ISLs, we still require the LoS condition:
\begin{align}
    \psi\leq\arccos{\frac{r_e}{r_e+h_l}}+\arccos{\frac{r_e}{r_e+h_m}}=\psi_{lm},\label{eq:leo-meo-los}
\end{align}
and the condition for the absolute elevation angle:
\begin{align}
    |\theta|\geq\theta_{lm}=25\text{\textdegree},\label{eq:leo-meo-theta}
\end{align}
where
\begin{align}
    \theta=\arctan\Big({\frac{\cos\psi-(r_e+h_l)/(r_e+h_m)}{\sin\psi}}\Big).\label{eq:leo-meo-theta2}
\end{align}
Clearly, Eq. \eqref{eq:leo-meo-theta2} is equivalent to
\begin{align}
    \Big|{\frac{\cos\psi-(r_e+h_l)/(r_e+h_m)}{\sin\psi}}\Big|\geq\tan{\theta_{lm}},\label{eq:leo-meo-theta3}
\end{align}
or the range for $\psi$ from solving Eq. \eqref{eq:leo-meo-theta3}:
\begin{align}
    \psi\leq\psi_{lm}^+\text{ or }\psi\geq\psi_{lm}^-,
\end{align}
where $\psi_{lm}^+$ and $\psi_{lm}^-$ are the solutions when the LHS of Eq. \eqref{eq:leo-meo-theta3} is equal to $\pm\tan{\theta_{lm}}$, respectively. Unlike user-to-satellite communication, LEO-to-MEO communication requires that both Eqs. \eqref{eq:leo-meo-los}-\eqref{eq:leo-meo-theta2} hold. We illustrate the concepts in satellite geometry in Fig. \ref{fig:elevation}.

\subsection{The Walker-Delta Constellation}

\begin{figure}[t]
\centering
\includegraphics[width=0.48\textwidth]{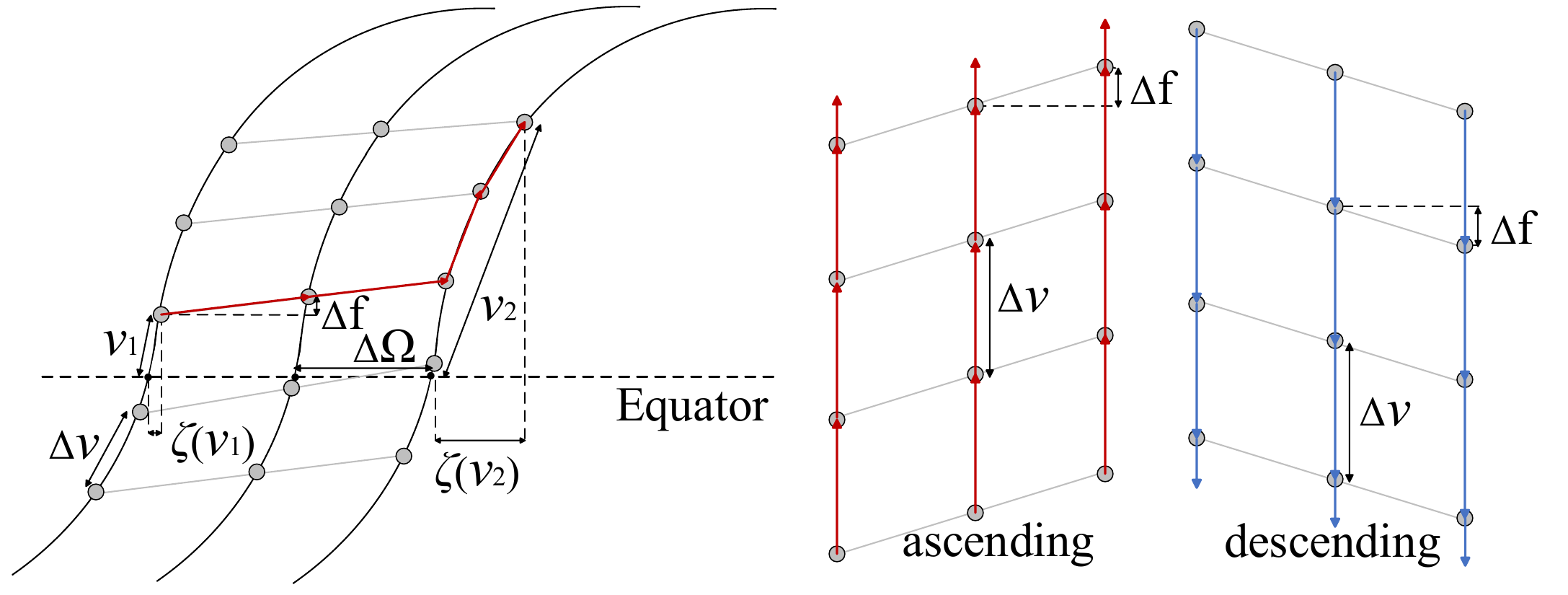}
\vspace{-5pt}
\caption{Left: The satellite constellation and and path with $H_p=2$, $H_q=2$. Right: The rotational directions of ascending and descending satellites.}
\vspace{-10pt}
\label{fig:constellation}
\end{figure}

Suppose that a satellite constellation consists of $P$ orbit planes and $Q$ satellites per orbit plane, where all orbit planes have the same inclination $\alpha$, all $P$ ascending nodes are equally spaced along the Equatorial plane, and all $Q$ satellites are equally distributed around the orbit, as depicted in Fig. \ref{fig:constellation}. Clearly, the RAAN difference between adjacent orbit planes is $\Delta\Omega=2\pi/P\!\in\![0,2\pi)$, and the phase offset between adjacent satellites in the same orbit is $\Delta\nu=2\pi/Q\!\in\![0,2\pi)$. The phase offset between adjacent orbit planes $\Delta f=2\pi F/(PQ)$ specifies the difference in the true anomaly $\nu$ between two horizontal neighbors, where $0\!\leq\!F\!<\!P$ is the phasing factor. In this way, $P\!\cdot\!\Delta\nu$ is always an integer multiple of $\Delta\nu$, a critical requirement to ensure the regularity of satellite spacing; otherwise, the relative angular offset between satellites in adjacent planes would not repeat among orbit planes, leading to non-periodic phasing patterns. In the literature, such a satellite constellation is formally denoted by $\alpha:PQ/P/F$. Currently, the massive satellite mega-constellation might comprise multiple constellations with varying inclination $\alpha$ and height $h$, in which each constellation is called a shell. For example, Starlink has 6 shells as of the year 2025 \cite{starlinkfiling}.

There are currently two commercial options of the parameters, i.e., Walker-Star (medium $\alpha$) and Walker-Delta (high $\alpha$). In essence, the choice marks a design trade-off between coverage for the mid-latitudes with Walker-Delta (e.g., OneWeb \cite{oneweb}, Iridium \cite{iridium}) versus global coverage including the pole areas with Walker-Star (e.g., GPS \cite{gps}, Galileo \cite{galileo}, Starlink \cite{starlink}, Kuiper \cite{kuiper}). In this paper, we consider Walker-Star for the MEO controllers and Walker-Delta for the LEO switches, respectively.

\subsection{Satellite Grid (Torus), Inter-Satellite Link, and Hop Count}

In Walker-Delta constellations, we denote the $q$-th satellite in the $p$-th orbit plane with an index $(p, q)$. The Keplerian coordinates at time $t$ are
\begin{align}
    \Omega_t(p,q)=&\ \mathcal{N}(p\cdot\Delta\Omega-\omega_e\cdot t),\label{eq:Omega_t}\\
    \nu_t(p,q)=&\ \mathcal{N}(p\cdot\Delta f+q\cdot\Delta\nu+\omega_s\cdot t)\label{eq:nu}
\end{align}
for all $0\leq p<P$ and $0\leq q<Q$.

Except in polar areas, each satellite establishes four stable ISLs with its neighbors, i.e., two intra-plane peers indexed by $(p, (q - 1)\text{ mod }Q)$ and $(p, (q + 1)\text{ mod }Q)$, and two inter-plane peers indexed by $((p - 1)\text{ mod }P, q)$ and $((p + 1)\text{ mod }P, q)$. Locally, the constellation exhibits a grid-like topology where a satellite has ISL connectivity with four adjacent peers \cite{starlinkfiling}. Globally, the constellation form a torus where satellites wrap around at the boundary $p=P$ and $q=Q$. As a result, any pair of satellites are connected by one or more of these ISL hops. It is worth emphasizing that ISLs are not established between adjacent ascending and descending satellites, as their proximity is transient and incidental rather than sustained.

Currently, the majority of research works (e.g., \cite{dynamics1,dynamics2}) adopt the criterion of hop count instead of the sum of LoS distances to select the shortest path. For laser links, the per-hop processing time often outweighs the propagation delay, rendering a shorter aggregate link length insufficient to offset the cost of extra hops. At time $t$, for a pair of satellites indexed by $(p_1,q_1)$ and $(p_2,q_2)$, we can derive a general formula for the minimum hop count
\begin{align}\label{eq:H}
    H=H_p+H_q.
\end{align}
In Eq. \eqref{eq:H}, $H_p$ denotes the inter-plane hop count and $H_q$ denotes the intra-plane hop count:
\begin{align}
    H_p=&\ \min\{|p_1-p_2|, P-|p_1-p_2|\}, \label{eq:H_p}\\
    H_q=&\ \min\{|q_1-q_2|, Q-|q_1-q_2|\}, \label{eq:H_q}
\end{align}
which is obtained from the torus-like property, where a path takes either the forward or the reverse direction. In addition, we can assume that each terrestrial user always has access to at least one ascending satellite and one descending satellite \cite{dynamics1}. Then communication between two terrestrial users can be implemented through four link types: ascending–ascending (A2A), ascending–descending (A2D), descending–descending (D2A), and descending–ascending (D2D).

\section{Segment Routing Background} \label{sec:sr-background}

SR \cite{sr} is a source-routing paradigm that encodes a path as an ordered list of segment identifiers (SIDs) in the IPv6 segment routing header (SRH) at ingress nodes. This allows only the ingress node to maintain SR policies, thereby simplifying TE, improving scalability and robustness, and reducing the burden on the control plane.

\subsection{The SR Control Plane and Data Plane}

The SR architecture is specifically designed with the flexibility of the hardware infrastructure, either to leverage existing multi-protocol label switching (MPLS) network (SR-MPLS) with IGP-extensions and provide backward compatibility for a smoother migration, or to be built natively on an ubiquitous IPv6 network (SRv6) and unlock advanced network programming capabilities \cite{srsurvey}. There exist core differences between SR-MPLS and SRv6 in the data plane, for which SR-MPLS employs MPLS labels as segment identifiers (SIDs) and encodes the route as a label stack, whereas SRv6 utilizes 128-bit IPv6 addresses as SIDs and carries the SL through the segment routing header (SRH). Similar distinctions appear in the control plane, where SR-MPLS still requires coordinating a globally consistent segment routing global block (SRGB) to allocate SIDs, whereas SRv6 completely eliminates label coordination by using IPv6 address prefixes as SIDs and can summarize SIDs like regular IPv6 routes.

\subsection{Segment Routing Domains}

\begin{figure}[H]
    \centering
    \includegraphics[width=.48\textwidth]{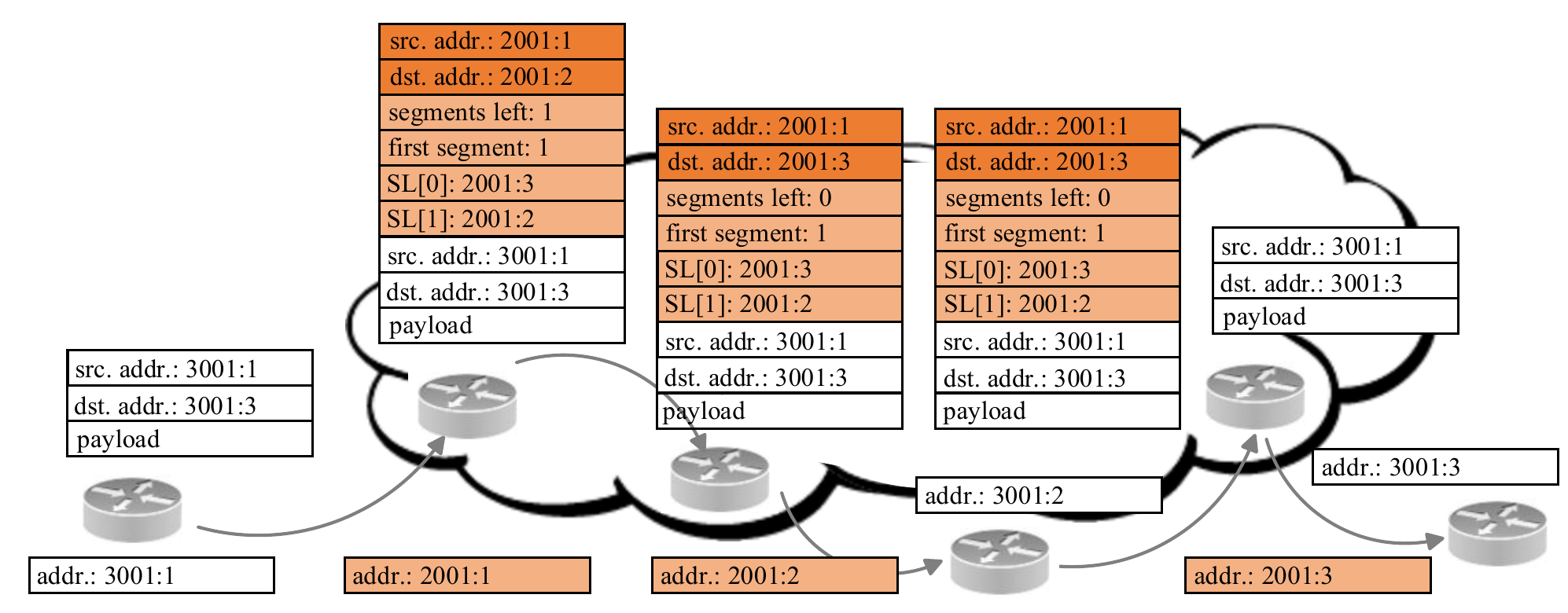}
    \vspace{-10pt}
    \caption{SRD 2001 consists of SR-capable nodes 2001:1-2001:3 and the IP domain consists of SR-incapable nodes 3001:1-3001:3. Each packet diagram denotes the packet header when emitted through the egress edge. The fields include the outer/encapsulated header (in dark orange), the SRH (in orange), and the payload (including inner header, etc.), resp., with the active segment always appearing in the outer header.}
    \label{fig:SRH}
\end{figure}

Since the Internet is expanding, segment routing domains (SRDs) \cite{ToN17} have been introduced primarily for scalability, administrative control, and security isolation. Informally, an SRD consists of a number of SR-capable nodes, and the routing details within the SRD can be hidden from the outside. To this end, for instance, SRDs use binding SIDs (BSIDs) that enable boundary nodes to handle intra-domain routing, expanding the BSID into complete path that can adapts to internal changes. Hybrid SR/IP networks exist in the gradual hardware migration process, in which SR-capable nodes interoperate with legacy IP nodes, but have significant implications for SRD topologies: SR-MPLS requires each SRD form a connected subnet since IP nodes will discard packets, while SRv6 lifts this restriction by allowing IP nodes to simply ignore the SRH and forward packets using legacy IGP rules \cite{ToN20}. Consequently, SR-capable nodes can be deployed dispersedly with SRv6 \cite{ToN20}, although fragmented SRDs remain administratively undesirable \cite{sr}.
In fully SR or hybrid networks that support multiple disjoint SRDs, every SRv6 node programs its SIDs into the forwarding information base (FIB). In addition, the SRD boundary nodes (i.e., ingress and egress nodes) assume greater responsibility than the interior nodes. For instance, ingress nodes map incoming traffic flows to SLs through SR policies, while egress nodes perform SRH removal and packet decapsulation \cite{mo2}. Fig. \ref{fig:SRH} illustrates a typical path in a hybrid IPv6 network, with SR-capable nodes 2001:1–2001:3 within SRD 2001.

SR permits multiple traversals of the same SRD to improve TE performance by leveraging external paths \cite{ToN17}. However, intra-SRD flows are discouraged from taking external paths due to latency, security, and QoS concerns. While multiple traversals are allowed, policy nesting is prohibited to avoid routing loops and segment list depth (SLD) violations \cite{mo1}. Consequently, we forbid nested IPv6 headers and nested SLs, requiring packet decapsulation upon exiting the current SRD. In Section \ref{sec:satellitedynamics}, we establish four path modes, which all have the minimum hop counts and are thus loop-free.

\section{System Overview} \label{sec:systemoverview}

To address the aforementioned challenges, we propose a framework that jointly addresses domain partitioning, SR-TE, and BSID translation. Concretely, the workflow consists of three phases:

1) Domain Partitioning Module. This module gathers information about MEO/LEO satellites and ISL states and packages it into standardized message formats. During the VT snapshot $\tau$ (about 15 seconds for Starlink), the network topology is considered static. This module aims to partition the constellation into disjoint SRDs, to minimize the control overhead in terms of the total time required for flow table update and domain state synchronization.

2) SR-TE Module. This module solves the problem of minimizing network congestion under the gravity model for traffic matrices. In particular, it leverages the recently proposed D-PDLP library to obtain communication-efficiency and multi-processor parallelism without performance degradation. One benefit of the distributed solution is that the routing variables are localized within their corresponding SRDs, thus preserving SRD-level privacy. Another benefit is that it suffices to communicate the variables in $\mathrm{AllReduce}$ (summation) format, thus significantly reducing communication overhead. 

3) SL Translation Module. This module translates the resulting routing variables into actionable SLs.

\section{MEO Satellite Control Domain Partitioning}

This section introduces the hierarchical MEO/LEO satellite constellations model and outlines the optimization objectives for control domain partitioning. During a VT snapshot interval $\tau$, the hierarchy consists of MEO satellites as upper-level control nodes and LEO satellites as lower-level switching nodes. MEO satellites have computational capabilities to perform route computation and network management in the control plane, while LEO satellites function as packet forwarding switches in the data plane, maintaining and updating flow table entries in the dedicated TCAM hardware. We modify the control overhead model proposed by a related work \cite{eunomia}.

\subsection{Persistence of User-to-Satellite and LEO-to-MEO FoVs}

To reduce unnecessarily frequent handovers, we prefer user-to-satellite and LEO-to-MEO connectivity that are persistence over a long duration. Due to LEO satellites' predictable positions over time, an MEO satellite can use either two-line elements (TLEs) or stored ephemeris to predict future LEO satellite positions in its FoV. However, verifying that the FoV condition holds for continuous time typically requires dense sampling or numerical root finders, both of which are computationally expensive. Under mild and practical conditions, we propose an algorithm that ensures persistent connectivity throughout the VT snapshot interval $\tau$ by checking the FoV conditions only at the start time $t_1$ and the end time $t_2$. This algorithm is formalized in Lemma 1 and Lemma 2, with proofs provided in Appendix A.

\vspace{10pt}

\noindent\textit{Lemma 1 (Persistence of User-to-Satellite FoVs)}. Define $n=\sqrt{\mu_E/r_a^3}$ as the satellite's mean motion and $\mu_E$ as the Earth's gravitational field strength. Suppose that the current VT snapshot starts at $t_1$ and ends at $t_2$ with duration $\tau=t_2-t_1$, and the geocentric angles are $\psi_1$ and $\psi_2$. If $\psi_1\leq\psi_{us}$, $\psi_2\leq\psi_{us}$, and $\psi_1$ and $\psi_2$ have sufficient margins:
\begin{align}\label{eq:condition1}
    |f(\psi_1)-f(\psi_{us})|+|f(\psi_2)-f(\psi_{us})|\geq\tau,
\end{align}
then we can ensure $\psi_t\leq\psi_{us}$ for all $t\in[t_1,t_2]$. The definite integral $f$ is defined as:
\begin{align*}
    f(\psi)=\int_c^\psi\frac{\sqrt{D-2k\cos u\sin u\sqrt{D}+k^2\sin^2u}}{n\sin^2 u}du
\end{align*}
for any constant $c$, with $k=r_e/r_a$, and $D$ is defined as:
\begin{align*}
    D=1-2k\cos u+k^2.
\end{align*}
The condition in Eq. \eqref{eq:condition1} is sufficient but not necessary, which we can verify efficiently with numerical integration Python libraries such as $\mathrm{SciPy}$ \cite{scipy}.

\vspace{10pt}

\noindent\textit{Lemma 2 (Persistence of LEO-to-MEO FoVs)}. Let $n_l$ and $n_m$ be the mean motions of the LEO and MEO satellites. If $\psi_1\leq\psi_{lm}$, $\psi_2\leq\psi_{lm}$, $\psi_1\leq\psi_{lm}^+$, $\psi_2\leq\psi_{lm}^+$, and $\psi_1$ and $\psi_2$ have sufficient margins for the LoS condition:
\begin{align}\label{eq:condition2}
    |g(\psi_1)-g(\psi_{lm})|+|g(\psi_2)-g(\psi_{lm})|\geq\tau,
\end{align}
and have sufficient margins for the elevation angle condition:
\begin{align}\label{eq:condition3}
    |g(\psi_1)-g(\psi_{lm}^+)|+|g(\psi_2)-g(\psi_{lm}^+)|\geq\tau,
\end{align}
where the definite integral $g$ is defined as:
\begin{align*}
    g(\psi)=\int_c^\psi\frac{\sqrt{D-2k\cos u\sin u\sqrt{D}+k^2\sin^2u}\sqrt{D}r_m}{(n_lr_l+n_mr_m)\sin^2 u}du
\end{align*}
for any constant $c$, with $k=r_l/r_m$, and $D$ is defined as:
\begin{align*}
    D=1-2k\cos u+k^2.
\end{align*}
Then we can ensure $\psi_t\leq\psi_{lm}$, $\psi_t\leq\psi_{lm}^+$ for all $t\in[t_1,t_2]$. Similar FoV conditions apply for negative elevation angles, namely $\psi_1\leq\psi_{lm}$, $\psi_2\leq\psi_{lm}$, $\psi_1\geq\psi_{lm}^-$, and $\psi_2\geq\psi_{lm}^-$. However, they cannot be simultaneously satisfied for the combination of Starlink and GPS constellations. Like for user-to-satellite FoVs, the conditions in Eqs. \eqref{eq:condition2}-\eqref{eq:condition3} are sufficient but not necessary.

\vspace{10pt}


\begin{figure}[t]
    \centering
    \includegraphics[width=.48\textwidth]{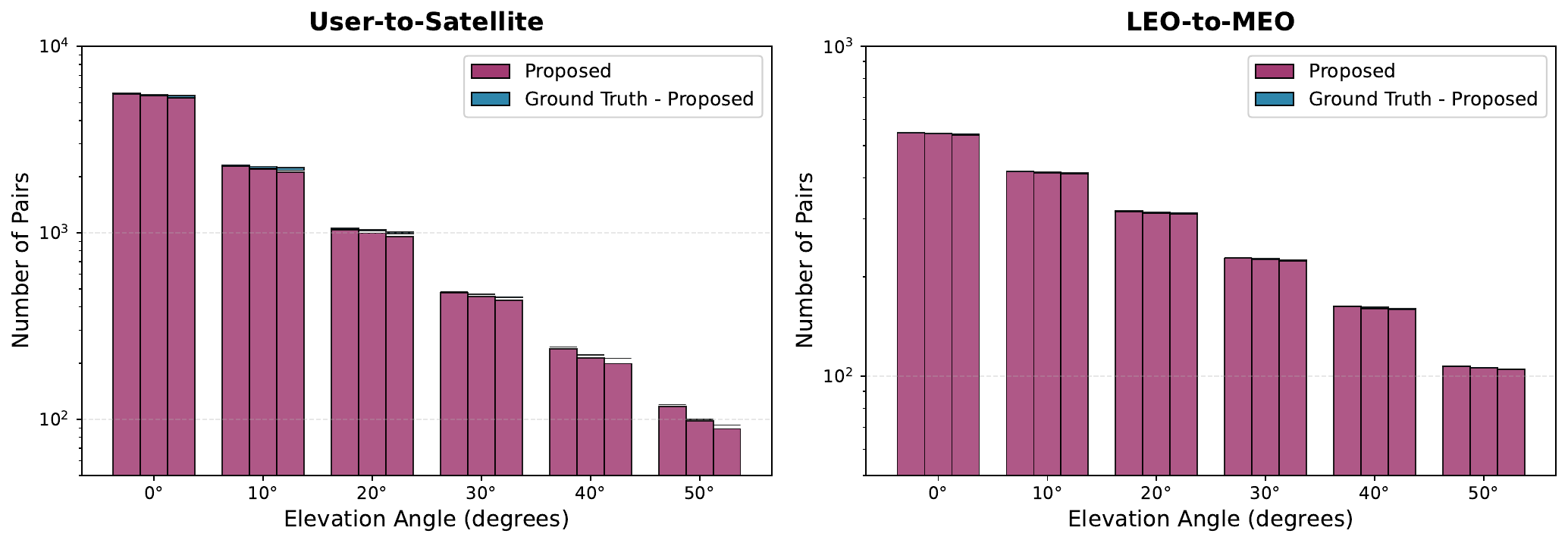}
    \vspace{-10pt}
    \caption{The x‑axis shows the elevation angle, with three bars plotted at each angle corresponding to time intervals $\tau=10$, 20, and 30 seconds. The y‑axis, displayed on a logarithmic scale, indicates the number of matched pairs.}
    \vspace{-10pt}
    \label{fig:estimation}
\end{figure}

While these lemmas provide sufficient conditions that tend to underestimate, experiments show that they generate results that are pretty close to the ground truth. For the Starlink constellation $53.0\text{\textdegree}:1584/72/3/550$ km and the GPS constellation $55.0\text{\textdegree}:24/6/4/20,200$ km, setting a practical margin of $15\text{\textdegree}$ for $\psi$ yields $\tau\sim123$ s for Lemma 1 and $\tau\sim69$ s for Lemma 2 by performing straightforward calculation. Figure \ref{fig:estimation} presents a comparative analysis of methods for estimation of ground-to-Starlink and Starlink-to-GPS pairs. The numerical methods proposed in Lemma 1 and Lemma 2 consistently perform well for all minimum required elevation angles ranging from $0\text{\textdegree}$ to $50\text{\textdegree}$, with the closest results occurring in the user-to-satellite settings, particularly at practical, medium minimum elevation angles. Despite the difference, the estimation method only slightly underestimate the matching pair counts compared with the ground truth.

\subsection{Control Overhead Model}\label{sec:upperlevel}

During VT snapshot $N$, the total control overhead $T^\mathrm{ctrl}$ for all MEO controllers depends on domain partition and consists of three main components. That is,
\begin{align*}
    \textstyle\max_iT_i^\mathrm{sync} + \max_iT_i^\mathrm{migr} + \sum_iT_i^\mathrm{pdhg} + \max_iT_i^\mathrm{flow},
\end{align*}
which represent the overhead of global state synchronization, partial variable migration, aggregate variable communication, and flow table updates, respectively. The explicit dependence on current VT snapshot is omitted when possible. We list key notations used in this section in Table \ref{tab:domainpartition}.

\begin{table}[ht]
    \centering
    \caption{Notations Frequently Used in the control overhead model}
    \label{tab:domainpartition}
    \begin{tabular}{m{1.5cm} m{6.5cm}}
        \hline
        \textbf{symbol} & \textbf{description} \\
        \vspace{1pt}
        $\tau$ & VT snapshot duration \\
        \vspace{1pt}
        $N, N\!-\!1$ & indices of current, previous VT snapshot \\
        \vspace{1pt}
        $\mathrm{SRD}_i$ & $i$-th SRD \\
        \vspace{1pt}
        $\mathrm{FoV}_i$ & FoV of $i$-th MEO controller \\
        \vspace{1pt}
        $M_{\mathrm{sync}}$ & LEO-to-MEO synchronization message size \\
        \vspace{1pt}
        $M_{\mathrm{comm}}$ & MEO-to-MEO communication message size \\
        \vspace{1pt}
        $M_{\mathrm{flow}}$ & MEO-to-LEO flow table update messages size \\
        \vspace{1pt}
        $S_{j}$ & $\#$ synchronization messages to send state variables for $j$-th LEO switch \\
        \vspace{1pt}
        $G_{j}$ & $\#$ synchronization messages to migrate PDGH solver variables for $j$-th LEO switch \\
        \vspace{1pt}
        $N_{j}$ & $\#$ flow update messages for $j$-th LEO switch \\
        \vspace{1pt}
        $V$ & $\#$ communication messages to transmit aggregate (i.e., $\mathrm{AllReduce}$) variables \\
        \vspace{1pt}
        $B_{ji}$ & bandwidth from $j$-th LEO switch to $i$-th MEO controller \\
        \vspace{1pt}
        $B_{ii^{\prime}}$ & bandwidth from $i$-th MEO controller to $i^{\prime}$-th MEO switch \\
        \vspace{1pt}
        $B_{ij}$ & bandwidth from $i$-th MEO controller to $j$-th LEO switch \\
        \vspace{1pt}
        $d_{ji}^{\mathrm{prop}}$ & propagation delay from $j$-th LEO switch to $i$-th LEO controller \\
        \vspace{1pt}
        $d_{ii^{\prime}}^{\mathrm{prop}}$ & propagation delay from $i$-th to $i^{\prime}$-th MEO controller \\
        \vspace{1pt}
        $d_{ij}^{\mathrm{prop}}$ & propagation delay from $i$-th MEO controller to $j$-th LEO switch \\
        \vspace{1pt}
        $N^{\mathrm{pdhg}}$ & $\#$ PDHG updates during $\tau$ \\
        \hline
    \end{tabular}
\end{table}

\subsubsection{LEO-to-LEO Global State Synchronization Overhead} Since control domain partition $\mathrm{SRD}_i\subset\mathrm{FoV}_i$ is not known yet, all LEO switches within $\mathrm{FoV}_i$ have to send state variables (e.g., $B_{ij}$ and $d_{ij}^{\mathrm{prop}}$) to the MEO controller sequentially. The synchronization overhead for $i$-th MEO controller is:
\begin{align*}
T_i^{\mathrm{sync}} = \textstyle\sum_{j\in\mathrm{FoV}_i} \displaystyle\Big( \frac{S_j \cdot M_{\mathrm{sync}}}{B_{ji}} + d_{ji}^{\mathrm{prop}} \Big).
\end{align*}
Note that $T_i^{\mathrm{sync}}$ does not affect the optimization of partition $\mathrm{SRD}_i$, since the sum must be taken over all $j\in\mathrm{FoV}_i$, which is determined by verifying the FoV conditions in Lemma 2 for each pair of $(i,j)$. 

\subsubsection{MEO-to-MEO Partial Variable Migration Overhead} Due to satellite mobility, $i$-th MEO controller must migrate PDHG solver variables corresponding to $\mathrm{SRD}_i$ to an adjacent MEO. The migration overhead for $i$-th MEO controller is:
\begin{align*}
    T_i^{\mathrm{migr}}=\textstyle\sum_{j\in\mathrm{SRD}_i^{N\!-\!1}}\sum_{i^{\prime}\not=i}b_j^{i^{\prime}}\displaystyle\Big(\frac{G_j\cdot{M}_{\mathrm{sync}}}{B_{ii^{\prime}}}+d_{ii^{\prime}}^{\mathrm{prop}}\Big).
\end{align*}
In the migration process, $i$-th MEO controller hands over solver variables corresponding to LEO switches that were part of $\mathrm{SRD}_i^{N\!-\!1}$, i.e., those under its control in the preceding VT snapshot, but are reassigned to a different $\mathrm{SRD}_{i^{\prime}}$ with $i^{\prime}\not=i$ in current VT snapshot.

\subsubsection{MEO-to-MEO Aggregate Variable Communication Overhead} For each PDHG update iteration, the aggregate variables in $\mathrm{AllReduce}$ format are expected to traverse all MEO controllers. The total communication overhead is:
\begin{align*}
    T_i^\mathrm{pdhg} = N^{\mathrm{pdhg}}\cdot\textstyle\sum_{i}\displaystyle\Big( \frac{V \cdot M_{\mathrm{comm}}}{B_{ii^{\prime}}} + d_{ii^{\prime}}^{\mathrm{prop}} \Big).
\end{align*}
The PDHG scheme in Section \ref{sec:lpformulation} requires that the aggregate variables traverse all MEO controllers exactly once in arbitrary order. Since the size of aggregate variables $V$ to transmit from $i$-th to $i^{\prime}$-th controller is constant and thus irrespective of partition $\mathrm{SRD}_i$, by treating the summand in $T_i^{\mathrm{pdhg}}$ as the cost of travel from $i$-th to $i^{\prime}$-th controller, we can convert it exactly into the NP-hard traveling salesman problem (TSP), whose optimization depends solely on the MEO topology and has no bearing on domain partition $\mathrm{SRD}_i$.

\subsubsection{MEO-to-LEO Flow Table Update Overhead} When the MILP is solved, $i$-th MEO controller transmits the solution to flow split ratios back to all LEO switches $j\in\mathrm{SRD}_i$, and the LEO switches update the corresponding flow table entries in the local TCAM. The flow table update overhead for $i$-th MEO controller is:
\begin{align*}
T_i^{\mathrm{flow}} = \textstyle\sum_{j\in\mathrm{SRD}_i} \displaystyle\Big( \frac{N_j\cdot M_{\mathrm{flow}}}{B_{ij}} + d_{ij}^{\mathrm{prop}} \Big).
\end{align*}

\subsubsection{Other Overhead} The related work \cite{eunomia} also introduces path computing overhead, which is subsumed into the PDHG solver's runtime and thus falls outside the scope of this work.

\begin{figure}[t]
\centering
\includegraphics[width=0.48\textwidth]{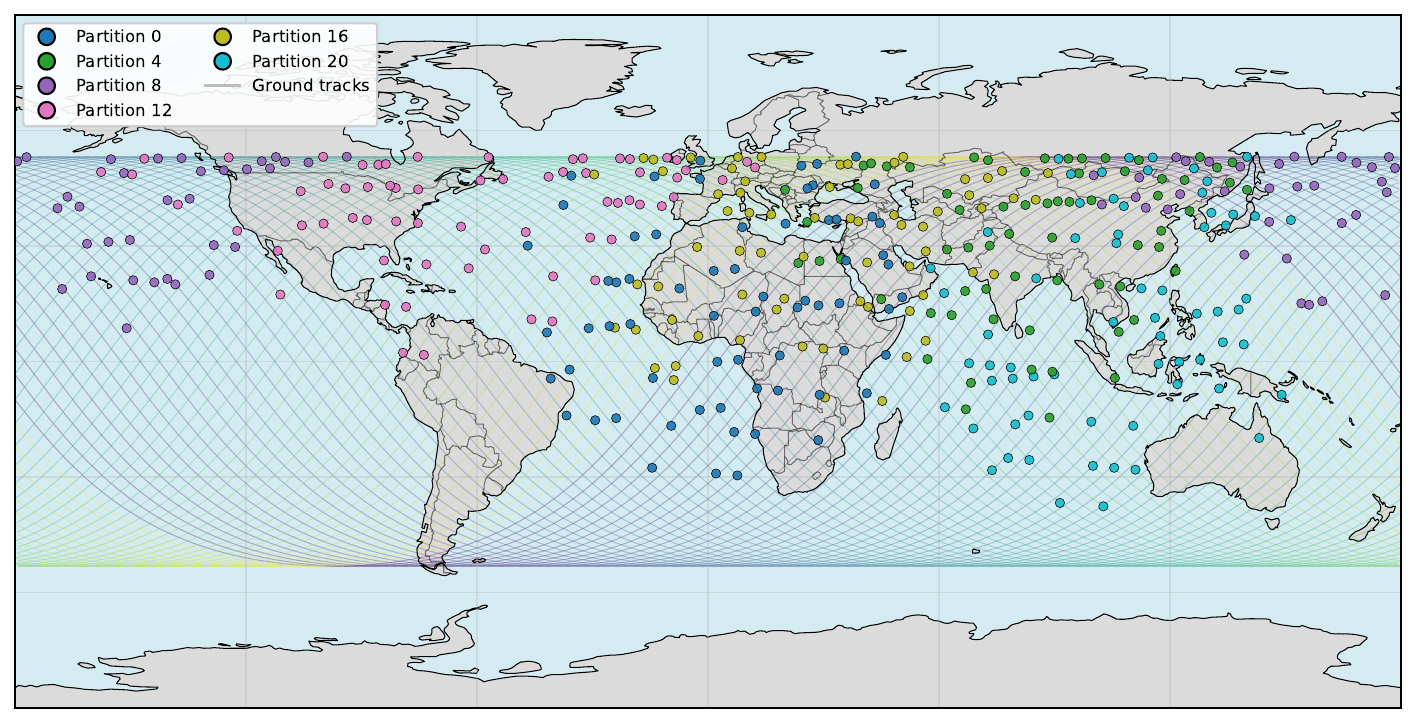}
\vspace{-10pt}
\caption{Starlink domain partitions (0, 4, 8, 12, 16, 20) shown on a global map. Each partition has a distinct color; satellites as colored circles, ground tracks as faint lines. Land: light gray; oceans: light blue. Partitions are geographically dispersed across all longitudes.}
\vspace{-10pt}
\label{fig:partitions}
\end{figure}

\subsection{Problem Formulation with MILP}

We assume $M_\mathrm{sync}=24$ Bytes \cite{eunomia}, $M_\mathrm{flow}=36$ Bytes \cite{eunomia}, $B_{ij}=20$ Gbps, and $G_j$ and $N_j$ depend on the solution of the TE problem for VT snapshot $N\!-\!1$. For any VT snapshot $N$, we can solve an equivalent MILP to minimize the total control and communication overhead:
\begin{align}
    \textstyle\min_{b_j^i}&\big(\max_iT_i^{\mathrm{migr}}+\max_iT_i^{\mathrm{flow}}\big) \label{eq:objective}\\
    \text{s.t.:}&\ b_{j}^{i}\leq\mathbb{I}[j\in\mathrm{FoV}(i)],\ \forall j,\ \forall i\label{eq:const1}\\
    &\ \!\textstyle\sum_{i}b_{j}^{i}=1,\ \forall j,\label{eq:const2}\\
    &\ b_{j}^{i}\in \{0,1\},\ \forall j,\ \forall i.\label{eq:const3}
\end{align}
The binary indicator variable is defined as $b_{j}^{i}=\mathbb{I}[j\in\mathrm{SRD}_i]$. The constraints in Eqs. \eqref{eq:const1}-\eqref{eq:const3} require that an LEO switch belongs exactly one MEO controller, and it must reside within the FoV of that particular controller. Figure \ref{fig:partitions} illustrates an example partition obtained by solving the MILP for Starlink and GPS with $\mathrm{gurobipy}$ \cite{gurobi}.

\section{Segment Routing Traffic Engineering Subproblem}\label{sec:sr-te}

Analogous to persistent LEO-to-MEO matching in the domain partition problem, we seek persistent user-to-satellite matching in the SR‑TE problem. Given stable matching within a VT snapshot, the problem can be formulated as an LP.

\subsection{Terrestrial User Assignment}\label{sec:terrestrialuserassignment}

With the persistence of FoV conditions in Lemma 1, we can associate a terrestrial user with a satellite in its proximity. We prioritize satellites with a higher elevation angle $\theta$, under the constraint that the user is within the satellite's FoV during the VT snapshot interval $\tau$. Prioritizing satellites with a higher elevation angle is justified by several factors for physical efficiency, such as reduced free-space path loss, lower atmospheric attenuation, decreased multipath interference, improved phased-array antenna gain, and longer dwell times leading to fewer handovers. To this end, we need to solve two separate problems of the form
\begin{align}
    \textstyle\max_{p,q}&\ \theta\Big(\big(\phi_u,\lambda_u,0\big),\big(\phi_t(p,q),\lambda_t(p,q),h\big)\Big),\label{eq:elevation}\\
    \text{s.t.:}&\ \text{user $u$ is persistently in sat. $(p,q)$'s FoV during $\tau$},\notag
\end{align}
one for $\nu_t\in[-\pi/2,\pi/2)$ and the other for $\nu_t\in[-\pi,-\pi/2)\cup[\pi/2,\pi)$, resp. For each terrestrial user at $(\phi_u,\lambda_u,0)$, we solve the two problems and find the two associated satellites indexed by $(p,q)$, one ascending and the other descending.

\subsection{Randomized Shortest Path Construction}

Then, the hop count between two terrestrial users depends on four path modes $\text{X}2\text{X}$ with hop counts $H^\text{X2X}$, where $\text{X}\in\{\text{A},\text{D}\}$. Unlike most works that consider only the minimum hop count across all four path modes, we reserve all modes to facilitate load balancing and enhance fail-safe robustness. For any source-destination user pair, we can associate the source and destination with an ascending and a descending satellite, respectively, thereby establishing four distinct paths. Each path takes the minimum horizontal and vertical hop counts in Eqs. \eqref{eq:H_p}-\eqref{eq:H_q}, restricting the path within the spherical rectangle. We adapt the probabilistic approach \cite{dynamics2}, in which each source satellite randomly determines whether to change direction at each step with random coinflips having success rate $p$. Once the path has completed all required horizontal hops, for example, it is forced to complete the path with all the remaining vertical hops, and vice versa. A similar approach, one-turn rectilinear routing \cite{one-turn}, has been adopted in wireless networks, which with appropriately designed turning rules can indeed reduce the maximum link utilization across the entire network.

\begin{figure}[t]
\centering
\includegraphics[width=0.245\textwidth]{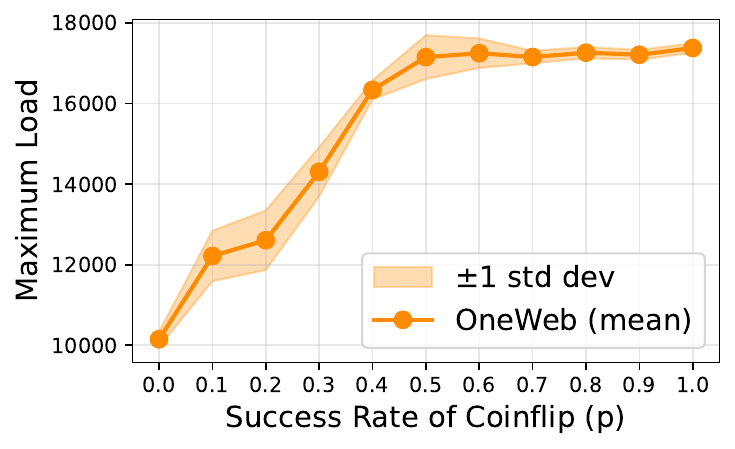}
\includegraphics[width=0.235\textwidth]{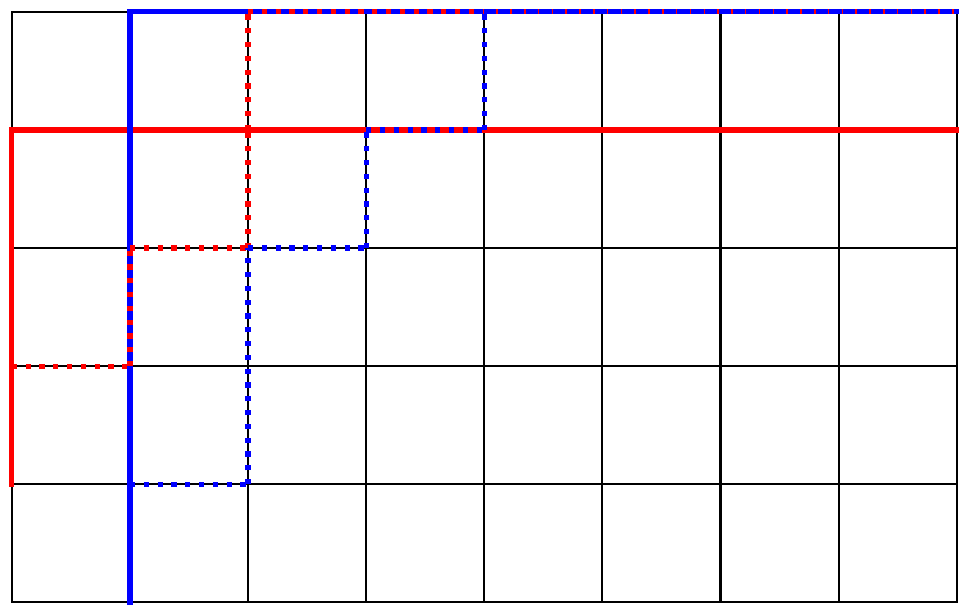}
\vspace{-20pt}
\caption{Impact of success rate $p$ on maximum load. Left: Maximum load as a function of $p$, showing a monotonic increase that saturates for $p\geq0.5$. Right: Illustration of the underlying cause: paths with $p=0$ (solid arrow) are edge-disjoint, whereas paths with $p>0$ (dashed lines) potentially exhibit edge overlaps, leading to increased contention and higher maximum load.}
\vspace{-10pt}
\label{fig:coinflips}
\end{figure}

For a given source-destination pair, we observe that while numerous minimum-hop-count paths exist, the path corresponding to success rate $p=0$, which prioritizes all horizontal hops before any vertical hops, consistently yields the lowest maximum load for sparse pairs. This trend is illustrated in Fig. \ref{fig:coinflips}, which depicts the impact of the success rate $p$ on the maximum load. The left subplot demonstrates that the maximum load increases with $p$ and plateaus for $p\geq0.5$. The right subplot offers insight into this behavior: the $p=0$ path (solid arrow) exhibits no overlapping edges, whereas paths with $p>0$ (dashed lines) potentially share multiple edges, leading to contention and consequently higher maximum load. In this work, we consistently require $p=0$.

\vspace{-10pt}

\begin{table}[ht]
    \centering
    \caption{Notations Frequently Used in the SR-TE}
    \label{tab:lowerlevel}
    \begin{tabular}{m{1.7cm} m{6.3cm}}
        \hline
        \textbf{symbol} & \textbf{description} \\
        \vspace{1pt}
        ${G}({V}, {E})$ & network ${G}$ with nodes ${V}$ and directed edges ${E}$ \\
        \vspace{1pt}
        $i\in{V}$ & LEO satellite $i$ \\
        \vspace{1pt}
        $e\in{E}$ & directed ISL link $e$ \\
        \vspace{1pt}
        $s, d$ & terrestrial source/destination user \\
        \vspace{1pt}
        $U$ & network congestion (i.e., maximum link utilization) \\
        \vspace{1pt}
        $C({e}) \geq 0$ & capacity of link $e$ \\
        \vspace{1pt}
        $t^{sd} \geq 0$ & traffic demand from $s$ to $d$ \\
        \vspace{1pt}
        $T$ & traffic matrix where the $(s,d)$-th entry is $t^{sd}$\\
        \vspace{1pt}
        $\text{path}^{sd}_\text{XX}$ & path from $s_\text{X}$ to $d_\text{X}$, where $\text{X}\in\{\text{A},\text{D}\}$ \\
        \vspace{1pt}
        $f^{sd}_\text{XX}\in[0,1]$ & fraction of $t^{sd}$ routed over $\text{path}^{sd}_\text{XX}$, where $\text{X}\in\{\text{A},\text{D}\}$ \\
        \vspace{1pt}
        $t^{sd}_\text{XX} \geq 0$ & traffic of $t^{sd}$ routed over $\text{path}^{sd}_\text{XX}$, where $\text{X}\in\{\text{A},\text{D}\}$ \\
        \vspace{1pt}
        $t^{sd}_\text{XX}(e) \geq 0$ & traffic of $t^{sd}_\text{XX}$ traversing link $e$, where $\text{X}\in\{\text{A},\text{D}\}$ \\
        \vspace{1pt}
        $\text{SRD}_k$ & $k$-th SRD, $\forall k\in K$ \\
        \hline
        \vspace{1pt}
    \end{tabular}
\end{table}

\subsection{Problem Formulation with LP}\label{sec:lpformulation}

The LP formulation for the SR-TE problem is straightforward, attempting to minimize the maximum ISL utilization over the LEO satellite constellation using split ratios \cite{splitratios}:
\begin{align}
    \min&\ U \label{eq:lpa}\\
    \text{s.t.:}&\ f^{sd}_\text{AA}+f^{sd}_\text{AD}+f^{sd}_\text{DA}+f^{sd}_\text{DD}=1,\ \forall s,d\in V \label{eq:lpb}\\
    & \textstyle\sum_{s,d}t^{sd}_\text{AA}(e)+t^{sd}_\text{AD}(e)+t^{sd}_\text{DA}(e)+t^{sd}_\text{DD}(e) \leq U \cdot C(e),\notag \\
    &\ \qquad\qquad\qquad\qquad\qquad\quad\quad\quad\quad\ \forall e\in E \label{eq:lpc} \\
    &\ t^{sd}_\text{AA}(e)=\mathbb{I}[e\in\text{path}^{sd}_\text{AA}]\cdot f^{sd}_\text{AA}\cdot t^{sd},\ \forall s,d\in V \label{eq:lpd} \\
    &\ t^{sd}_\text{AD}(e)=\mathbb{I}[e\in\text{path}^{sd}_\text{AD}]\cdot f^{sd}_\text{AD}\cdot t^{sd},\ \forall s,d\in V \label{eq:lpe} \\
    &\ t^{sd}_\text{DA}(e)=\mathbb{I}[e\in\text{path}^{sd}_\text{DA}]\cdot f^{sd}_\text{DA}\cdot t^{sd},\ \forall s,d\in V \label{eq:lpf} \\
    &\ t^{sd}_\text{DD}(e)=\mathbb{I}[e\in\text{path}^{sd}_\text{DD}]\cdot f^{sd}_\text{DD}\cdot t^{sd},\ \forall s,d\in V \label{eq:lpg} \\
    &\ U,\ f^{sd}_\text{AA},\ f^{sd}_\text{AD},\ f^{sd}_\text{DA},\ f^{sd}_\text{DD}\geq0,\ \forall s,d\in V. \label{eq:lph}
\end{align}
The notation $f^{sd}_\text{AA}$ refers to the traffic split ratio for the A2A path between $s,d$; the notations $f^{sd}_\text{AD}$, $f^{sd}_\text{DA}$, and $f^{sd}_\text{DD}$ are defined analogously. We list all notations frequently used in the SR-TE in Table \ref{tab:lowerlevel}.

At any time, given the traffic demands $t^{sd}$ are known and the positions of terrestrial users $s$ and $d$, their associated satellites $s_\text{A}$, $s_\text{D}$, $d_\text{A}$ and $d_\text{D}$ are determined. Each path between a source–destination satellite pair is randomly constructed once, prior to execution, and reused for all traffic demands. The indicator variables are self-evident; for example, $\mathbb{I}[e\in\text{path}^{sd}_\text{AA}]$ equals 1 if the path from $s_\text{A}$ to $d_\text{A}$ goes through link $e$, and 0 otherwise. The problem has linear objective and linear constraints with variables given by Eq. \eqref{eq:lph}.

While the optimal solution can be obtained using highly efficient off‑the‑shelf LP solvers, such as Gurobi \cite{gurobi}, we instead propose a distributed LP solution. With this method, the LEO switches are partitioned into disjoint SRDs, and the MEO controller of each SRD plans the routing for the LEO switches internally. A notable benefit of this distributed design is that the routing decisions, i.e., the randomly chosen paths and split ratios, in one SRD are not disclosed to other SRDs, and only the aggregated $\mathrm{AllReduce}$ variables are shared.

\subsection{Distributed Solution with D-PDLP}

The primal–dual hybrid gradient (PDHG) framework has demonstrated great potential for solving large-scale LPs. Under this umbrella, methods based on PDHG for LP (PDLP) \cite{pdhg} have established the state-of-the-art for solving massive-scale LPs. To date, distributed PDLP (D-PDLP) \cite{d-pdlp} has been proposed that scales across multiple GPUs.

\subsubsection{PDLP} PDLP-based methods consider the general LP
\begin{align}\label{eq:generalprimal}
    \textstyle\min_{x\in\mathcal{X}}c^Tx\text{ subject to }Ax\in\mathcal{S}
\end{align}
where $A\!\in\!\mathbb{R}^{m\times n},c\!\in\!\mathbb{R}^n$, feasible set $\mathcal{X}=\{x\in\mathbb{R}^m:l_v\!\leq\!x\!\leq\!u_v\}$, constraint range $\mathcal{S}=\{s\in\mathbb{R}^n:l_c\!\leq\!s\!\leq\!u_c\}$, and all upper/lower bounds can extend to infinity. The corresponding Lagrangian leads to an equivalent saddle-point formulation:
\begin{align}\label{eq:generaldual}
    \textstyle\min_{x\in\mathcal{X}}\max_{y\in\mathcal{Y}}c^Tx+y^TAx-p(-y;l_c,u_c),
\end{align}
where the piece-wise linear function $p(y;l,u)=u^Ty^+-l^Ty^-$ with $y^+=\max\{y,0\}$ and $y^-=\max\{-y,0\}$, and the dual feasible set $\mathcal{Y}\subset\mathbb{R}^m$ is the component-wise Cartesian product that depends on the boundedness of the primal
constraints $\mathcal{X}$:
\begin{align*}
    \mathcal{Y}_i = \begin{cases}
            \{0\}, & \text{if } (l_c)_i=-\infty, \ (u_c)_i=\infty, \\
            \mathbb{R}^-, & \text{if } (l_c)_i=-\infty, \ (u_c)_i<\infty, \\
            \mathbb{R}^+, & \text{if } (l_c)_i>-\infty, \ (u_c)_i=\infty, \\
            \mathbb{R}, & \text{otherwise.}
        \end{cases}
\end{align*}
The PDHG method can be applied to the saddle-point problem in Eq. \eqref{eq:generaldual}, yielding the following update rules \cite{pdhg}:
\begin{align*}
    x^{t+1}=&\ \mathrm{Proj}_{\mathcal{X}}\big(x^t-\lambda_1(c-A^Ty^t)\big),\\
    z^{t+1}=&\ A(2x^{t+1}-x^t),\\
    y^{t+1}=&\ y^t-\lambda_2z^{t+1}-\lambda_2\mathrm{Proj}_{-\mathcal{S}}\big(\lambda_2^{-1}y^t-z^{t+1}\big)
\end{align*}
where $\lambda_1$ and $\lambda_2$ denote the primal and dual stepsizes, respectively, and $\mathrm{Proj}_{-\mathcal{S}}$ denotes the projection to the set $-\mathcal{S}$.

\subsubsection{D-PDLP} To scale across multiple distributed compute resources, D-PDLP
adopts a flexible 2D partitioning strategy to distribute vector $c$, matrix $A$, and variables $x,y,z$, allowing the solver to effectively benefit from data parallelism. We rewrite the LP in Eqs. \eqref{eq:lpa}-\eqref{eq:lph} in the form of Eq. \eqref{eq:generalprimal} and partition the corresponding matrix $A$ column-wise, creating a special case of the general 2D partitioning strategy,
\begin{align}\label{eq:d-generalprimal}
    \textstyle\min_{x\in\mathcal{X}}\sum_{k\in K}c_k^Tx_k\text{ subject to }\sum_{k\in K}A_kx_k\in\mathcal{S},
\end{align}
with $x=[U;x_{1}^T;\ldots;x_{K}^T]^T$, $c=[1;0_{1}^T;\ldots;0_{K}^T]^T$, $l_v=0$, $u_v=\infty$. For all $k\in K$,
\begin{align*}
    x_{k}=&\ \big[\{f^{sd}_\text{AA}\}_{s,d\in V,s_\text{A}\in\mathrm{SRD}_k}^T;\{f^{sd}_\text{AD}\}_{s,d\in V,s_\text{A}\in\mathrm{SRD}_k}^T;\\
    &\ \{f^{sd}_\text{DA}\}_{s,d\in V,s_\text{D}\in\mathrm{SRD}_k}^T;\{f^{sd}_\text{DD}\}_{s,d\in V,s_\text{D}\in\mathrm{SRD}_k}^T\big]^T.
\end{align*}
The all-zero vectors $0_k$ in $c$ have dimensions equal to their corresponding part $x_k$. The matrix $A$ and constraint range $\mathcal{S}$ are determined by Eqs. \eqref{eq:lpb}-\eqref{eq:lpc} and their exact expressions are omitted here.

To use the D-PDLP framework across multiple MEO controllers, we assign $x_k$, $c_k$, $A_k$, and the corresponding $y_k$, $z_k$ to $k$-th MEO controller, with $U$ assigned to any of them. The complete distributed procedure is outlined in Alg. \ref{alg:d-pdlp}, where the $\mathrm{AllReduce}$ operator aims to aggregate vectors across all devices and broadcast the sum back. Note that the traffic split ratios in $x_k^t$ are not shared across SRDs; only the aggregated variables $z^t$ are shared. This is critical for maintaining the confidentiality of management information within each SRD.

\begin{algorithm}
\caption{D-PDLP}
\label{alg:d-pdlp}
\begin{algorithmic}[1]
\renewcommand{\algorithmicrequire}{\textbf{Input:}}
\REQUIRE problem instance $(A,c,\mathcal{X},\mathcal{S})$, domains $\mathrm{SRD}_k$, initial value $(x^0,y^0)$, stepsizes $\lambda_1$ and $\lambda_2$, and accuracy $\epsilon$.
\STATE Determine $(A_k,c_k,\mathcal{X}_k)$ according to $\mathrm{SRD}_k$, $\forall k\in K$.
\FOR {$t = 1$ to $t_{\mathrm{max}}$}
\STATE $\triangleright$ Primal update
\STATE Compute partial gradient $g_k=A_k^Ty^t$.
\STATE Update primal variable $x_k^{t+1}\!=\!\mathrm{Proj}_{\mathcal{X}_k}(x_k^t\!-\!\lambda_1(c_k-g_k))$.
\STATE $\triangleright$ Distributed dual update
\STATE Compute local partial constraint $v_k=A_k(2x_k^{t+1}-x_k^t)$.
\STATE Aggregate $v_k$ across columns $z^{t+1}=\mathrm{AllReduce}(v_k)$.
\STATE Receive $z^{t+1}$ and update local dual variable in parallel $y^{t+1}=y^t-\lambda_2z^{t+1}-\lambda_2\mathrm{Proj}_{-\mathcal{S}}\big(\lambda_2^{-1}y^t-z^{t+1}\big)$.
\STATE Stop if the KKT error is less than $\epsilon$ \cite{pdhg,d-pdlp}.
\ENDFOR
\RETURN $x^t=\{x^t_k\}_{k\in K}$.
\end{algorithmic}
\end{algorithm}

The main computation overhead is the multiplication between the matrices and vectors in lines 5 and 7, which is distributed and performed in parallel by the MEO controllers. The main communication overhead comes from the $\mathrm{AllReduce}$ operation, i.e. the sum of vectors $v_k$, in line 8. The summation over distributed $v_k$ with $\mathrm{AllReduce}$ requires communication across over all $K$ MEO controllers. Fortunately, this can be achieved using two parallel traversals via the bidirectional Ring-AllReduce procedure in a distributed manner. The computation in line 9 is performed identically at each device to save communication overhead.

A notable caveat is that the convergence of PDHG-based methods is generally slow owing to their first-order nature. To mitigate this, we propose a hybrid strategy that employs a faster commercial solver, such as Gurobi \cite{gurobi}, to generate an initial solution, followed by potential invocations of D-PDLP to dynamically refine the solution in response to fluctuations observed during the VT snapshot interval.

\section{Performance Evaluation}\label{sec:evaluation}

For evaluations, we construct MEO-LEO satellite networks by importing known specifications (e.g., topologies, ISL capacities, and geodesic distance between LEO satellites' terrestrial nadirs) and traffic matrices $T$ through synthetic and sampling techniques.
\vspace{-5pt}

\subsection{Hardware and Software Setup}

All experiments were conducted on a dedicated workstation with an Intel i9-14900KF/RTX 4090 GPU workstation with 32 GB RAM, using Ubuntu 20.04-LTS and Python 3.11.14. We obtain an initial solution of the LP with $\mathrm{gurobipy}$ \cite{gurobi}, and apply distributed adjustment of the initial solution to traffic demand fluctuations with D-PDLP \cite{d-pdlp} when necessary. 
\vspace{-5pt}

\subsection{Simulation Setup}

We implement the astrodynamic models in Section \ref{sec:satellitedynamics} with Python, and formulate and solve the MILP in Section \ref{sec:upperlevel} and LP in Section \ref{sec:lpformulation} with $\mathrm{gurobipy}$ \cite{gurobi}.

\subsubsection{LEO Constellations} We conduct extensive experiments on commercial LEO constellations as the LEO switches:
\begin{itemize}
    \item Starlink S1: $53.0\text{\textdegree}:1584/72/3/550$ km,
    \item Kuiper K1: $51.9\text{\textdegree}:1156/34/3/630$ km, and
    \item OneWeb Phase 1: $87.9\text{\textdegree}:588/12/3/1,200$ km.
\end{itemize}

\subsubsection{MEO Constellation} We use the baseline GPS constellation $55.0\text{\textdegree}:24/6/4/20,200$ km as the MEO controllers. The satellites are arranged in six orbital planes with four primary satellites in each plane, which ensure global coverage.

\subsubsection{Geographic Datasets and Traffic Matrices}

We obtain city-level geographic information systems (GIS) geographic datasets from GeoNames \cite{geonames}, a free and reliable foundation for collecting demographic (e.g., population) and geographic (e.g., latitude, longitude) data of the cities worldwide. We use the gravity model \cite{gravity} that assumes the traffic demand $t^{sd}$ between nodes $s$ and $d$ is proportional to the product of their respective capacities $C^s,C^d$ and inversely proportional to some function of the distance $\text{dist}(s,d)^{\kappa}$ between them, i.e.,
\begin{align}\label{eq:gravity}
    \bar{t}^{sd}=\frac{C^s\cdot C^d}{\mathrm{dist}(s,d)^\kappa}.
\end{align}
We define $C^s$ as the total population of the city $s$, and denote $\mathrm{dist}(s,d)$ as the minimum hop count between the city pair $(s,d)$. Because the maximum hop count is small and constant for all city pairs, we can disregard the dependence on distance between city pairs, i.e., $\kappa=0$. To account for the variations across time, we can treat traffic demands as Poisson random processes, whose means are given by $\bar{t}^{sd}$ in Eq. \eqref{eq:gravity}. We independently sample each traffic demand from the Poisson distribution $t^{sd}\sim\mathrm{Poisson}(\bar{t}^{sd})$ and repeat this procedure.

\subsection{Baselines and Shared Hyperparameters}

Although a substantial body of current work focuses on routing in LEO constellations, research on minimizing link congestion remains relatively limited. To ensure a fair comparison, we select several baselines that operate at the same level of granularity—namely, the flow level—since packet-level information is prohibitively large and difficult to collect, update, and disseminate. Specifically, we compare our method against these baselines with respect to the primary objective (i.e., link utilization) as well as derived metrics, including end-to-end delay and packet drop rate. The selected baselines comprise classic MHCR-based algorithms for satellite networks:
\begin{itemize}
\item DisCoRoute \cite{dynamics2}, which arranges ISLs near polar regions to minimize the total link length,
\item FROG \cite{mhcr2}, where terrestrial users consider both visible ascending and descending satellites that yield the minimum hop count,
\end{itemize}
as well as load-balancing algorithms:
\begin{itemize}
\item VLB, a classic traffic-oblivious random routing method.
\end{itemize}

We filter out cities with population less than 4 million and calculate traffic demands according to Eq. \eqref{eq:gravity}, where The population variable is expressed in millions and the result is measured in Mbps. The LEO switches for Starlink, Kuiper, and OneWeb are assumed to have processing rates of 20, 40, and 60 Gbps, respectively, and an input buffer capacity of 10 packets. These processing rates are below the ISL bitrates, ensuring that the links do not constitute the system bottleneck.

\subsection{Performance Analysis}

\begin{figure*}[t]
\centering
\includegraphics[width=0.98\textwidth]{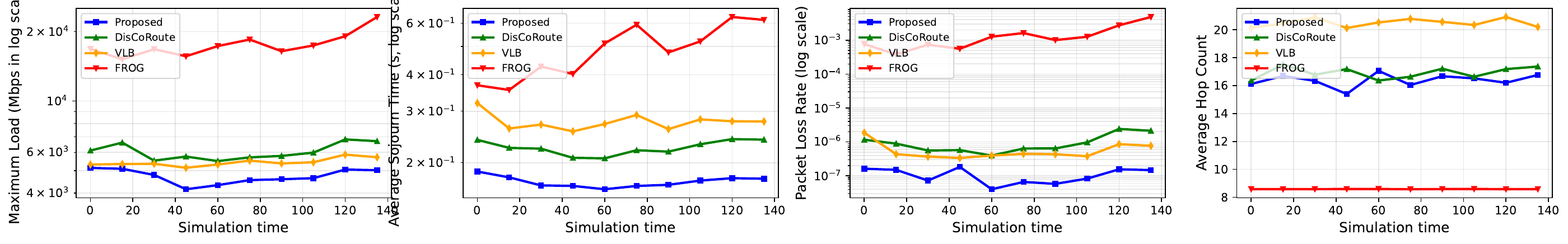}
\includegraphics[width=0.98\textwidth]{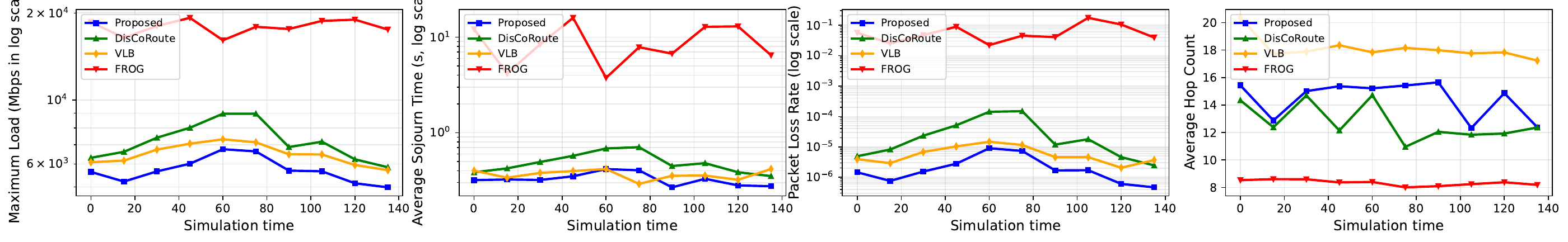}
\includegraphics[width=0.98\textwidth]{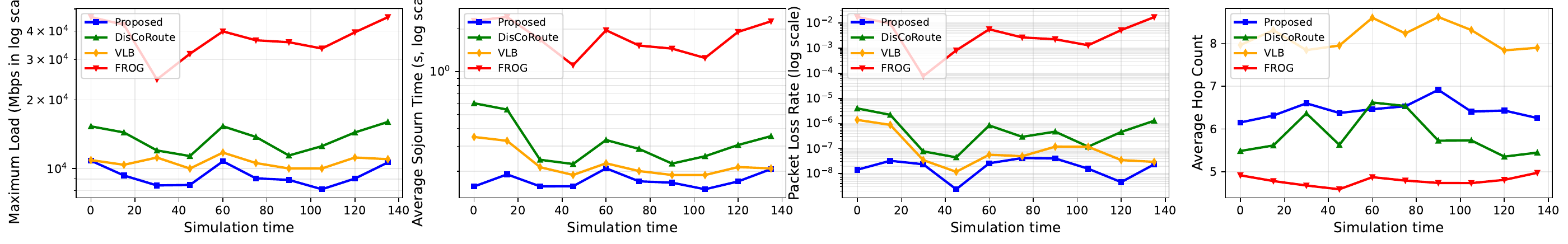}
\vspace{-10pt}
\caption{Performance comparison of the proposed algorithm against DisCoRoute, VLB, and FROG across four metrics. First row: Starlink, second row: Kuiper, third row: OneWeb. First column: maximum load (Mbps in log scale), second column: average sojourn time (s in log scale), third column: packet loss rate (log scale), fourth column: average hop count.}
\vspace{-10pt}
\label{fig:performance}
\end{figure*}

Fig. \ref{fig:performance} compares the performance of the proposed algorithm against DisCoRoute, VLB, and FROG across four metrics in Starlink, Kuiper, and OneWeb constellations. We can see from the subplots that our proposed algorithm excel in maximum load, average sojourn time, and packet drop rate, while being the second best in hop count. However, the proposed algorithm's non-optimality in hop count is a deliberate trade-off for superior load balancing. The simulation times simply denote the time of the LEO constellations, and there are not logical orderings in time.

\subsubsection{Maximum Load}

As shown in column one, the proposed algorithm achieves the lowest maximum load among all evaluated methods. In contrast, FROG exhibits greater loads at later time steps, indicating severe congestion under identical traffic conditions. DisCoRoute and VLB maintain intermediate maximum loads in between. Notably, these results validate that minimizing maximum load, rather than minimizing hop count, effectively prevents the formation of bottleneck links.

\subsubsection{Average Sojourn Time}

Column two presents the average sojourn times, which is the total time spent at the LEO switch including queueing delay and processing delay, across algorithms. The proposed method consistently achieves the lowest sojourn times, DisCoRoute and VLB exhibit moderate sojourn times, while FROG incurs the highest delays, especially at later time steps. This result is particularly noteworthy: despite not explicitly minimizing hop count, the proposed algorithm achieves superior delay performance by avoiding congested links. Conversely, FROG's strategy of minimizing hop count leads to the overutilization of specific links, resulting in queue buildup and increased per-packet waiting times. Thus, load balancing indirectly improves delay performance, even when path length is not the primary optimization objective.

\subsubsection{Packet Loss Rate}

The packet loss rates, shown on a logarithmic scale in column three, further reinforce the advantages of load balancing. The proposed algorithm maintains the lowest loss rates, DisCoRoute and VLB exhibit moderate loss rates, while FROG suffers from dramatically higher losses, reaching up to the order of $10^{-2}$. This disparity arises because hop-minimizing routes concentrate traffic onto a subset of links, causing buffer overflows and subsequent packet drops. The load-aware routing of the proposed algorithm distributes traffic more evenly across the network, reducing the probability that any single link reaches its capacity limit. These results underscore that loss minimization is a natural consequence of effective load balancing.

\subsubsection{Average Hop Count}

Column four reports the average hop count for each algorithm. FROG achieves the lowest hop count, confirming its primary design objective of minimizing path length. DisCoRoute and the proposed algorithm exhibit moderate hop counts, while VLB incurs the highest hop count. Critically, the proposed algorithm does not achieve the minimum hop count; it sacrifices path length optimality to attain better load distribution. This trade-off is deliberate and central to our design philosophy: in congested networks, a longer but less loaded path often yields lower delays, fewer losses, and higher overall throughput than a shorter but congested path. The superior performance of the proposed algorithm on all three primary metrics (maximum load, sojourn time, and loss rate) justifies this design choice. Thus, the non-optimality of hop count should be interpreted not as a limitation but as the necessary cost of achieving effective load balancing in dense LEO constellations.

\subsubsection{Comparison Across Constellations}

In comparing the constellations, we observe that link status exhibits more pronounced variability in sparsely connected topologies. Such configurations demand higher-capacity ISLs to attain performance metrics comparable to those of more densely connected counterparts. This suggests that constellation topology plays a critical role in determining not only link status variability but also the capacity requirements necessary to sustain consistent network performance.

\section{Related Works}\label{sec:relatedworks}

Due to increasing importance in satellite networks, a number of approaches
have been proposed to improve the efficiency and scalability of routing. In this section, we briefly review satellite routing methods that leverage spherical torus topology.

\subsection{Minimum Hop Count Routing (MHCR) Methods}

Traditional shortest-path-based routing protocols, e.g., open shortest path first (OSPF), struggle to cope with the frequent topological changes inherent to LEO satellite constellations. To address this limitation, proactive routing algorithms based on the OLSR protocol \cite{top1} have been proposed, which periodically exchange control messages with sizes dynamically adjusted to network conditions. Other studies have exploited the regular and predictable nature of LEO constellation dynamics to enable low-complexity routing, as exemplified by OPSPF \cite{top2}, ASER \cite{top3}, and RFC 9717 \cite{rfc9717}. For satellite routing in particular, a growing body of current research favors MHCR, since hop count rather than physical link length should serve as the primary routing metric in satellite networks \cite{mhcr1, dynamics2}. Within the MHCR paradigm, several simple yet insightful algorithms have been introduced. For instance, DisCoRoute \cite{dynamics2} shortens inter-plane inter-satellite links by placing them near the polar regions. The FROG algorithm \cite{mhcr2}, instead of always selecting the nearest satellite, chooses the first and last satellites from a candidate subset, which can yield substantial hop count reductions when both ascending and descending satellites are available. In addition, recent efforts have also explored additional link opportunities beyond the conventional four-neighbor assumption in dense constellations \cite{top4, top5, top6, top7}.

\subsection{Packet-Level Methods to Minimize Delay}

In contrast to flow-level models that primarily focus on minimizing utilization, packet-level approaches employ queueing-theoretic tools to make routing decisions based on the real-time queue status of servers. For example, the LSP strategy \cite{lb1} considers the load of the current satellite, ELB \cite{lb2} considers the load of the next-hop satellite, while TLR \cite{lb3} accounts for both and enables multi-path routing. A common limitation of these approaches, however, is that frequent state updates can trigger large-scale network-wide flooding. A recent approach \cite{srd2} mitigates this by employing geographical heuristics to forward packets between clusters via terrestrial relays, reducing the reliance on network-wide flooding. In summary, existing methods either require the dissemination of update information across the network or lack a global view of network status due to reliance on strictly local information.

\subsection{Zone-Based Satellite Routing}

As modern LEO satellite constellations continue to expand, zone-based hierarchical routing has emerged as a promising approach to enable efficient and scalable TE, and has attracted growing research attention. Current research has explored dynamic zoning mechanisms based on various criteria, including geographic location and connectivity \cite{zone1}, traffic intensity and link quality \cite{zone2}, and desired routing performance \cite{zone3}. A key advantage of this paradigm lies in its natural alignment with SRDs when SR is adopted. Within this framework, the LEO satellite constellation is partitioned into disjoint SRDs, and satellites use intra-domain and inter-domain segment routing to realize multi-path forwarding \cite{srd1, srd2, srd3}. However, current inter-domain routing methods predominantly rely on local topological proximity, leading to sub-optimal performance. Although recent advances have begun incorporating load-balancing mechanisms, many existing approaches still do not consider global load distribution across available inter-domain paths.

\section{Conclusion and Future Work}\label{sec:conclusion}

This paper presents a segment routing framework for LEO satellite constellations that addresses the critical challenges of scalability, control overhead, and load balancing. By offloading all route computation, topology discovery, and control signaling to a higher-altitude MEO layer, we have demonstrated that resource-constrained LEO switches can operate efficiently as pure data-plane forwarding elements.

We have made fourfold key contributions. First, we establish sufficient persistence of FoV conditions for user-to-satellite and LEO-to-MEO connectivity over VT snapshot intervals (Lemma 1 and Lemma 2), enabling handover-free operation and stable SRD assignments. Second, we formulate the MEO control domain partitioning problem as an MILP that minimizes control overhead while respecting FoV constraints, producing disjoint SRDs that persist throughout each VT snapshot. Third, we develope a multipath SR-TE formulation that leverages four minimum-hop-count path types (A2A, A2D, D2A, D2D) to balance traffic and minimize maximum link utilization. This LP is solved in a distributed manner using the PDHG framework, preserving SRD-level privacy and requiring only $\mathrm{AllReduce communication}$. Fourth, we translate the optimal split ratios into hardware-feasible SLs.

Simulation results on practical LEO constellations (Starlink, Kuiper, OneWeb) with a GPS MEO control layer validate our approach. Compared to classic MHCR-based (DisCoRoute, FROG) and traffic-oblivious (VLB) baselines, our proposed method consistently achieves the lowest maximum link utilization, the lowest average sojourn times, and the lowest packet loss rates, albeit with a moderate hop count that deliberately trades path length optimality for load balancing.

Several avenues remain for future work. First, extending the distributed PDHG framework to handle dynamic topology changes within a VT snapshot would improve responsiveness to real-time traffic fluctuations; however, since PDHG only leverages first-order information, the convergence is typically slow. Second, incorporating additional optimization objectives, such as energy efficiency, latency fairness, or congestion-aware rerouting, could further enhance network performance. Third, experimental validation on hardware testbeds or orbital demonstration missions would provide practical insights into the feasibility of our approach under real-world conditions.



\end{document}